\documentclass[aps,pra,twocolumn,showpacs,superscriptaddress,preprintnumbers]{revtex4}
\usepackage{amsmath}
\usepackage{amssymb}
\usepackage{graphicx}

\begin{document}


\title{Quantum Key Distribution over Probabilistic Quantum Repeaters}

\author{Jeyran Amirloo}
\email[]{jamirloo@uwaterloo.ca}
\affiliation{Institute for Quantum Computing and Department of Electrical and Computer Engineering,
University of Waterloo, 200 University Ave. W., Waterloo, ON, Canada N2L 3G1}

\author{Mohsen Razavi}
\affiliation{Institute for Quantum Computing and Department of Electrical and Computer Engineering,
University of Waterloo, 200 University Ave. W., Waterloo, ON, Canada N2L 3G1}
\affiliation{Institute of Integrated Information Systems, School of Electronic and Electrical Engineering,
University of Leeds, Leeds, U.K. LS2 9JT.}

\author{A. Hamed Majedi}
\affiliation{Institute for Quantum Computing and Department of Electrical and Computer Engineering,
University of Waterloo, 200 University Ave. W., Waterloo, ON, Canada N2L 3G1}

\date{\today}

\begin{abstract}
A feasible route towards implementing long-distance quantum key distribution (QKD) systems relies on probabilistic schemes for entanglement distribution and swapping as proposed in the work of Duan, Lukin, Cirac, and Zoller (DLCZ) [Nature {\bf 414,} 413 (2001)]. Here, we calculate the conditional throughput and fidelity of entanglement for DLCZ quantum repeaters, by accounting for the DLCZ self-purification property, in the presence of multiple excitations in the ensemble memories as well as loss and other sources of inefficiency in the channel and measurement modules. We then use our results to find the generation rate of secure key bits for QKD systems that rely on DLCZ quantum repeaters. We compare the key generation rate per logical memory employed in the two cases of with and without a repeater node. We find the cross-over distance beyond which the repeater system outperforms the non-repeater one. That provides us with the optimum inter-node distancing in quantum repeater systems. We also find the optimal excitation probability at which the QKD rate peaks. Such an optimum probability, in most regimes of interest, is insensitive to the total distance.

\end{abstract}


\maketitle

\section{Introduction}
Among many emerging applications offered by quantum information science, quantum key distribution (QKD) is the only one that has received commercial attention, \cite{Note_QKD}, and may soon be publicly available \cite{Chapuran09a}. The latter depends on our ability to reduce the cost of the system and to make it available, not only over short point-to-point links, but also over long-distance network connections. Long-distance quantum communication relies on quantum repeater systems, which, themselves, rely on a large number of quantum memory units with efficient coupling to light and long coherence times \cite{Briegel98a, Razavi09a, SPIE, OFC}. The original proposal for quantum repeaters by Briegel {\em et al.} relies on performing high-fidelity quantum operations for entanglement swapping and purification \cite{Briegel98a}. In their scheme, the requirements for implementing quantum repeaters are similar to those of a quantum computer. Nevertheless, recent progress in miniaturizing trapped-ion quantum systems \cite{ike_review} and in improving light-ion coupling \cite{sussex} has made the prospects of this approach more promising. In the meantime, and especially for QKD applications, there is an alternative approach to building quantum repeaters, which, instead of using deterministic gates for measurement and purification, relies on probabilistic operations and post-measurement purification. This approach, first proposed by Duan, Lukin, Cirac, and Zoller (termed DLCZ hereafter) for atomic-ensemble memories \cite{DLCZ}, is potentially simpler to implement and its underlying idea for entanglement distribution and swapping has been used and extended in numerous frameworks and proposals for quantum repeaters \cite{DLCZvariants}. In this paper, we analyze a single-hop DLCZ repeater system, by accounting for path, measurement, and coupling loss effects as well as the multiple-excitation effect in ensembles. The latter is a fundamental source of error for such systems, and it has been fully taken into account in our analysis. We find the generation rate of secure key bits for the DLCZ QKD protocol in both cases of with and without an intermediate repeater node. By comparing the two results, we obtain architectural insights into how such quantum repeaters must be designed as functions of their various system parameters.

One of the main features of the DLCZ-based protocols for entanglement distribution and connection is their ability to remove certain errors by post-measurement processing. These post measurements are commonly part of the application in hand, e.g., QKD, and not the entanglement generation scheme itself. As a result, the generated state at the end of the DLCZ entanglement-generation protocols is not necessarily highly entangled. This has been shown in theory \cite{Razavi06} and experiment \cite{kimble_DLCZ} by, respectively, calculating and measuring the fidelity and the concurrence of entangled states obtained via a single-hop DLCZ repeater. To evaluate the performance of such systems in practice, it is important to include the post-measurement effect in our analysis. This has been achieved in two ways in our paper. First, by using a general application-independent conditional measure, and second, by looking at the specific case of QKD. In the first approach, we look at the {\em conditional} fidelity and the rate of generating entangled states in the DLCZ repeaters when we virtually assume that the generated state is non-vacuum. The vacuum state is the typical erroneous outcome of the DLCZ repeater protocol, which can commonly be ruled out by post measurements. In the second approach, we employ the entangled states generated by the DLCZ entanglement distribution or repeater protocol in a QKD setup and find its secure key generation rate. The DLCZ QKD protocol effectively filters out most cases that reduce entanglement measures of the pre-measurement states.

Our QKD rate analysis for quantum repeaters addresses two important practical issues. First, we calculate the rate for a repeater setup that uses multiple quantum memories per node. Second, by using a normalized rate-per-memory measure, we include the cost factor in our analysis as quantum memories are the most precious constituents of the system. In a probabilistic setup such as DLCZ, an acceptable key generation rate can only be achieved if we employ a large number of memories in parallel. Moreover, to achieve the maximum rate, the system resources must be successively employed in the process of entanglement distribution and connection to successively generate entangled states for use in the QKD protocol. Razavi {\em et al.} have studied these issues in a generic quantum repeater setup, and, here, we employ their results in the specific case of memories with sufficiently long coherence times \cite{Razavi09a}.

We consider the original DLCZ protocol, with atomic ensembles as its quantum memories. We are not, however, restricted to using this particular type of memory, and, in fact, both the DLCZ scheme, and hence our analysis, can be applied to other types of memories that work on the basis of the collective enhancement of light-matter interaction \cite{DLCZ}. As mentioned earlier, such memories must be employed in large numbers and demonstrate long coherence times. Recently, coherence times in excess of 5 ms are demonstrated for cesium atoms in an atomic comb \cite{cohtime}. This is, in principle, sufficient to cover distances up to 1000 km, provided that a large number of logical memories can be employed in parallel \cite{Razavi09a}. Atomic ensembles can potentially be used as multiple logical memories by applying/collecting light at/from different directions.

The rest of this paper is organized as follows. In Sec.~\ref{Sec:SD} we describe the DLCZ protocols for entanglement distribution, entanglement swapping, and QKD. In Sec.~\ref{Sec:PA:ED} we first review the fidelity analysis given in \cite{Razavi06} for the DLCZ scheme for entanglement distribution and, then, extend it to the case of DLCZ quantum repeaters. We compare the two cases in terms of their effective fidelity and throughput---the rate at which entangled states are created---assuming that a large number of memories are being used in parallel. Section~\ref{Sec:PA:QKD} compares the two systems in terms of the generation rate of secure key bits in a QKD setup, and Sec.~\ref{Sec:Con} concludes the paper.

\section{System Description}
\label{Sec:SD}

\begin{figure}
\includegraphics [width=\linewidth]{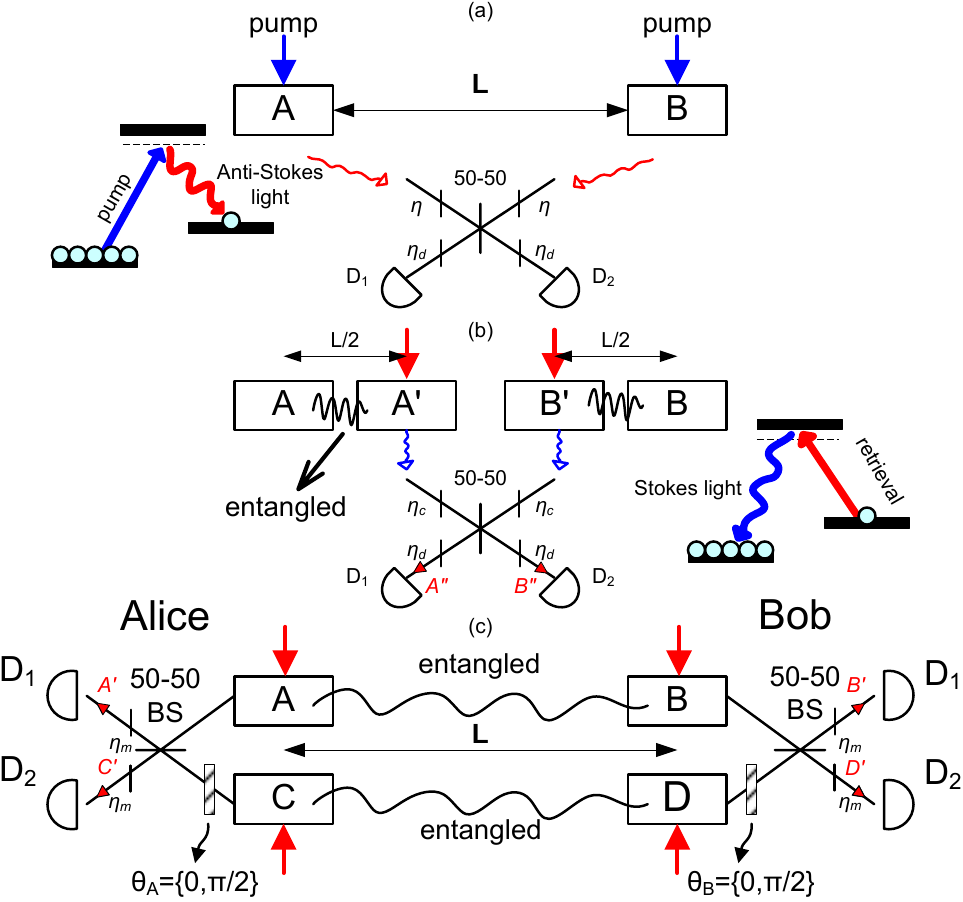}
\caption{(Color online) (a) The DLCZ scheme for entanglement distribution between atomic ensembles $A$ and $B$. The $\Lambda$-level atomic ensembles are coherently pumped such that the chance of driving more than one Raman transition in the two ensembles is low. The photons collected from such Raman transitions in a certain direction are then routed down to a midpoint, where a 50-50 beam splitter erases any which-way information. A single click on one, and only one, of photodetectors heralds entanglement between $A$ and $B$. (b) The DLCZ scheme for quantum repeaters. In order to entangle ensembles $A$ and $B$, at distance $L$, we first entangle $A\&A'$ and $B\&B'$, at distance $L/2$, using the scheme described in (a). We then use a 50-50 beam splitter and single-photon detectors to perform a partial Bell-state measurement on the photonic states retrieved from middle ensembles $A'$ and $B'$. A click on one, and only one, of detectors heralds the success of entanglement swapping. (c) The DLCZ setup for quantum key distribution. Alice and Bob first create two entangled pairs of ensembles, $A\&B$ and $C\&D$, using the DLCZ schemes described in (a) or (b). They perform QKD measurements by first converting the atomic states into photonic states, and then applying randomly chosen phase shifts, 0 or $\pi/2$, to the optical modes before they interfere at 50-50 beam splitters. That would effectively simulate an entanglement-based QKD protocol. A click on detector $D_{1/2}$ is associated with bit 1/0. In (a), the beam splitter with transmissivity $\eta$ represents the path loss. In (b), the beam splitter with transmissivity $\eta_c$ models the efficiency of the retrieval process. In (a) and (b), the beam splitter with transmissivity $\eta_d$ represents the quantum efficiency of photodetectors, and, in (c), the beam splitter with transmissivity $\eta_m = \eta_c \eta_d$ represents the total measurement efficiency. All schematically employed photodetectors in (a)--(c) have unity quantum efficiencies. Labels $A''$ and $B''$ in (b), and $A'$, $B'$, $C'$, and $D'$ in (c) represent the optical modes entering these ideal detectors. \label{SetupSchematic}}
\end{figure}

The DLCZ scheme for entanglement distribution works as follows; see Fig.~\ref{SetupSchematic}(a). Ensemble memories $A$ and $B$, at distance $L$, consist of atoms with $\Lambda$-level configurations, all initially in their ground states. By coherently pumping these atoms, some of them may undergo off-resonant Raman transitions that create anti-Stokes photons. The width and amplitude of the pump is chosen such that the probability of one such transition, $p_c$, is close to zero, hence the number of anti-Stokes photons, in the direction of interest, does not commonly exceed one. The resulting photons are routed down towards a 50-50 beam splitter located halfway between $A$ and $B$. The beam splitter erases any which-way information so that if, ideally, only one photon has been created at one of the ensembles, one and, at most, only one, of $D_1$ and $D_2$ clicks. According to the DLCZ protocol, if only detector $D_j$, $j=1,2$, in Fig.~\ref{SetupSchematic}(a), clicks, $A$ and $B$ are heralded to be ideally in the Bell state $|\psi_j^{\rm Bell}\rangle_{AB}= (|01\rangle_{AB} +(-1)^j|10\rangle_{AB})/\sqrt{2}$, where $|0\rangle_{K}$ is the ensemble ground state and $|1\rangle_{K} \equiv S_K^\dag |0\rangle_{K}$ is the symmetric collective excited state of ensemble $K=A,B$ with $S_K^\dag$ being the corresponding creation operator \cite{DLCZ,Razavi06}.

The fundamental source of error in the above DLCZ scheme is the multiple excitation effect, in which more than one anti-Stokes photon are created. Multiple photons passed through a lossy channel can reproduce an erroneous heralding event.  This effect can be alleviated, to some extent, by using photon-number resolving detectors (PNRDs), rather than non-resolving photodetectors (NRPDs). One click at a PNRD implies that exactly one photon is observed whereas, one click at an NRPD implies that {\em at least} one photon has been detected. In our forthcoming analysis, we fully consider the multiple-excitation effect when either type of detectors is employed and compare the system performance in various scenarios.

The 50-50 beam splitter together with the single photon detectors in Fig.~\ref{SetupSchematic}(a) effectively perform a partial Bell-sate measurement (BSM) on the incoming photons; see Fig.~\ref{BSM}. The DLCZ quantum repeater protocol uses this idea to distribute entanglement over longer distances. Figure~\ref{SetupSchematic}(b) shows the DLCZ repeater setup in which, we first entangle ensembles $A\&A'$ and $B\&B'$ using the DLCZ entanglement distribution protocol. We then perform a partial BSM on the retrieved photons from the middle ensembles $A'$ and $B'$, which, upon success, leaves $A$ and $B$ entangled.

One major application for memories entangled via the DLCZ schemes for entanglement distribution and repeater is the DLCZ QKD protocol. 
In this protocol, our two remote parties, Alice and Bob, first generate identical entangled pairs, namely $AB$ and $CD$, over distance $L$; see Fig.~\ref{SetupSchematic}(c). They then retrieve the photons in the four ensembles and perform a QKD measurement on these photons \cite{BB84}. The measurement modules used for this purpose is similar to the BSM module in Fig.~\ref{BSM} with additional phase shift units whose phase values are randomly picked to be either $0$ or $\pi/2$. These phase shifts are being applied to the photons retrieved from ensembles $C$ and $D$ in Fig.~\ref{SetupSchematic}(c). Alice and Bob repeat this experiment multiple times to create a raw key. After the sifting procedure, by which Alice and Bob specify the measurement events where they have both used the same phase shifts and have obtained at least one click on their respective detectors, they each obtain a sifted key by assigning bit one to their keys whenever only $D1$ has clicked on their side, and bit zero whenever only $D2$ has clicked. In the case of a double click, and only if NRPDs are being used, they assign bit zero or one, with equal probability, to their sifted keys. By using privacy amplification and reconciliation techniques, Alice and Bob turn their sifted keys to a secure key, which can be used for encryption purposes.

Throughout the paper, we assume that all setups in Fig.~\ref{SetupSchematic} are symmetric. In particular, we assume that the optical paths from ensembles to relevant detectors, in terms of accumulated phase and incurred loss, are identical. The retrieval efficiency, $\eta_c$, the quantum efficiency, $\eta_d$, and the measurement efficiency, $\eta_m = \eta_c \eta_d$ are also identical in all setups. To get the most out of our channel and detectors, we may need to use frequency up-converters or down-converters at the level of a single photon \cite{Albota06a}. We assume that the efficiency of such modules are also included in $\eta_m$ or path loss. In Fig.~\ref{SetupSchematic}, these loss effects are modeled by relevant beam splitters. All photodetectors in Fig.~\ref{SetupSchematic} have then unity quantum efficiencies. We furthermore assume that the dark current is negligible in all our photodetectors.

The achievable throughput for DLCZ protocols is commonly restricted by the probabilistic nature of its entanglement distribution and connection schemes. Let us consider the particular case of the DLCZ QKD protocol, which relies on two entangled pairs. The establishment of entanglement between $A$ and $B$, in Fig.~\ref{SetupSchematic}(c), is not necessarily coincident with the establishment of entanglement over $C$ and $D$. Hence if we use only two pairs of memories, we have to wait until we have two entangled pairs, and that reduces the rate. In order to get the most out of employed memories, we can employ a large number of logical memories in a parallel cyclic way as explained in \cite{Razavi09a}. By using a large number of memories at each site in parallel, we minimize the waiting time, and, therefore, maximize the rate. By using a cyclic protocol, we reuse memories as soon as they become available, and that increases the efficiency of our system. Throughout the paper, we assume the employment of a large number of memories per node. Given that each ensemble can be used as multiple logical memories, the number of physical systems required could be much fewer than the logical ones. By using a large number of memories, we also minimize the constraints on the coherence time of employed memories \cite{SPIE}. 

In what follows, we first review the performance of the DLCZ scheme for entanglement distribution reported in \cite{Razavi06}, and extend their results to the case of the DLCZ repeater. Then, we find the key generation rate for the DLCZ QKD protocol using entangled pairs created either directly by the DLCZ entanglement distribution scheme or by a single-hop repeater system. 

\section{Performance Analysis: Entanglement Distribution}
\label{Sec:PA:ED}
In this section, we first review the results reported in \cite{Razavi06} for a single DLCZ link, and then extend them to the case where one repeater node is used to create entanglement between the remote parties.

\subsection{DLCZ entanglement distribution: A review}
In \cite{Razavi06}, the joint state of $A$ and $B$ in Fig.~\ref{SetupSchematic}(a), $\rho_j^{AB}$, after a heralding event at $D_j$, $j=1,2$, is obtained. Here, we use the Fourier relation between a density operator and its anti-normally ordered characteristic function to rewrite $\rho_j^{AB}$ in the following form: 
\begin{eqnarray}\label{TransRhoChi}
\rho_j^{AB}=&{\displaystyle \int\frac{d^2\zeta_A}{\pi}\int\frac{d^2\zeta_B}{\pi}}&   \chi_A^{\rho_j^{AB} }(\zeta_A, \zeta_B) D_N(S_A,-\zeta_A)  \nonumber \\
&\!&  \times D_N(S_B,-\zeta_B) ,
\end{eqnarray}
where, for a complex variable $\zeta = \zeta_r+ i \zeta_i$ with $\zeta_r$ and $\zeta_i$ being real numbers, $\int {d^2\zeta} \equiv \int_{-\infty}^\infty{d {\zeta_r}} \int_{-\infty}^\infty{d {\zeta_i}}$, $D_N(a,\zeta) \equiv e^{\zeta a^\dag}e^{-\zeta^* a}$ is the normally ordered displacement operator for an annihilation operator $a$, and $\chi_A^{\rho_j^{AB} }$ is the anti-normally ordered characteristic function for $\rho_j^{AB}$. For a composite system of harmonic oscillators $A_1,\ldots,A_n$, with respective annihilations operators $a_1, \ldots, a_n$, the anti-normally ordered characteristic function for $\rho ^{A_1, \ldots, A_n}$, the joint state of the system, is defined as follows
\begin{equation}
\label{CharFuncDef}
\chi_A^{\rho^{A_1, \ldots, A_n}} (\zeta_1, \ldots,\zeta_n) \equiv \langle \prod_{i=1,\ldots,n} {D_A(a_i,\zeta_i)} \rangle,
\end{equation}
where $D_A(a,\zeta) \equiv e^{-\zeta^* a} e^{\zeta a^\dag}$ is the anti-normally ordered displacement operator for an annihilation operator $a$. We use a similar notation throughout the paper for relevant characteristic functions appearing in our analysis. Using Eqs.~(20)~and~(25) in \cite{Razavi06} along with Eq.~(\ref{CharFuncDef}), we obtain
\begin{eqnarray}
\label{CharFuncFirst}
\chi_A^{\rho^{AB}_j}(\zeta_A,\zeta_B)&=&
\exp[-\alpha(|\zeta_A|^2+|\zeta_B|^2)] \times \nonumber \\
&\!& (1-\frac{1}{2} \alpha|\zeta_A+(-1)^j\zeta_B|^2), \mbox{PNRD}
\end{eqnarray} 
and
\begin{eqnarray}
\label{CharFuncFirstN}
\chi_A^{\rho^{AB}_j}(\zeta_A,\zeta_B) &=& -\frac{1-p_c}{\eta_s^{AB}p_c}\exp[-\alpha(|\zeta_A|^2+|\zeta_B|^2)]\nonumber \\
&+& \exp\left[\frac{-\alpha\eta_s^{AB}p_c}{2(1-p_c)}|\zeta_A+(-1)^j\zeta_B|^2\right] \times \nonumber \\
&\,& \frac{\exp\left[-\alpha(|\zeta_A|^2 
+|\zeta_B|^2)\right]}{\alpha\eta_s^{AB}p_c}, \mbox{NRPD},
\end{eqnarray}
where $\alpha=\frac{1}{\eta_s^{AB}p_c+1-p_c}$ and $\eta_s^{AB} = \eta_d \eta$ is the total system efficiency. Here, $\eta = \exp(-(L_{AB}/2)/L_{att})$ represents the channel transmissivity in Fig. 1(a) with $L_{att}$ being the channel attenuation length and $L_{XY}$ denoting the distance between any two ensembles $X$ and $Y$. The main assumption in the above equations is that the employed setup is symmetric, i.e., the excitation probability $p_c$, the path loss and all relevant quantum efficiencies as well as incurring phase shifts are identical for all parties involved in the protocol.


For symmetric setups, the average fidelity of entanglement generated by the DLCZ entanglement distribution scheme is given by \cite{Razavi06}
\begin{eqnarray}\label{FidelityGeneral}
F^{(1)}_{AB} &= & \frac{1}{2}\sum_{j=1,2}\left._{AB}\langle\psi_j^{\rm Bell}|\rho^{AB}_j|\psi_j^{\rm Bell}\rangle_{AB}\right. \\
\label{FidelityNorepeater}
&=& \left\{\begin{array}{lr}(\eta_s^{AB}p_c+1-p_c)^3,&\mbox{PNRD}\\
(1-p_c)(\eta_s^{AB}p_c+1-p_c)^2,&\mbox{NRPD}\\\end{array}\right. .
\end{eqnarray}
The probability of heralding success for the above DLCZ entanglement distribution is given by \cite{Razavi06}
\begin{equation}\label{PjNorepeater}
P_{\rm herald}^{AB}=\displaystyle{\left\{\begin{array}{lr}\frac{\displaystyle 2(1-p_c)^2\eta_s^{AB} p_c}{\displaystyle (\eta_s^{AB} p_c+1-p_c)^3},&\mbox{PNRD}\\
\frac{\displaystyle 2(1-p_c)\eta_s^{AB} p_c}{\displaystyle (\eta_s^{AB} p_c+1-p_c)^2},&\mbox{NRPD}\\\end{array}\right.\equiv P_S(L_{AB}).}
\end{equation}

\begin{figure}
\includegraphics [width=1.5in]{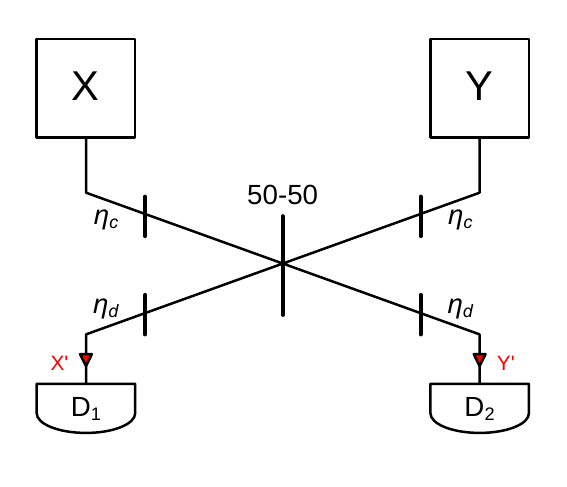}
\caption{(Color online) Partial Bell-state measurement (BSM) by linear optics. This is a common module in all DLCZ protocols in which the collective states of two atomic ensembles, $X$ and $Y$, are transferred to optical modes via a retrieval procedure \cite{DLCZ}. The resulting optical modes interfere at a 50-50 beam splitter before being detected by two single-photon detectors. In our BSM model, we denote the atomic-to-photonic conversion efficiency by $\eta_c$ and the photodetectors' quantum efficiencies by $\eta_d$. The single-photon detectors in the figure then have ideal unity quantum efficiencies. $X'$ and $Y'$ denote the optical modes entering these ideal photodetectors. \label{BSM}}
\end{figure}

\subsection{DLCZ Repeater Protocol}

The single-hop DLCZ repeater protocol works as follows. In order to create entanglement over distance $L$, the entire link is split into two segments of length $L/2$ as shown in Fig.~\ref{SetupSchematic}(b). Using the DLCZ protocol described in Fig.~\ref{SetupSchematic}(a), we first distribute entanglement between $A\&A'$ and $B\&B'$, and, then---only after we learn about the establishment of entanglement on both links---we perform a partial BSM on optical modes that are retrieved from ensembles $A'$ and $B'$; see Fig.~\ref{BSM}. 

In this section, we find the generation rate and the fidelity of entangled ensembles created over distance $L$ using the DLCZ quantum repeater of Fig.~\ref{SetupSchematic}(b) in different scenarios. The methodology we use here is similar to that of \cite{Razavi06} in that we first find the relevant characteristic functions for the state on which the BSM will be performed. Rate and fidelity can then be calculated in terms of integrals with Gaussian integrands, for which analytic results are obtained using the symbolic software Maple. Such results are commonly too lengthy to be presented in their explicit forms, and here we leave them in their compact integral forms. 

After the establishment of entanglement on the sublinks $AA'$ and $BB'$, the initial joint characteristic function of $AA'B'B$ is given by 
\begin{equation}\label{MultiplyCharRepeater}
\chi_A^{\rho^{AA'B'B}_{j,k}}(\zeta_A,\zeta_B,\zeta_C,\zeta_D)= \chi_A^{\rho^{AA'}_j}(\zeta_A,\zeta_B) \chi_A^{\rho^{B'B}_k}(\zeta_C,\zeta_D), 
\end{equation}
where $\rho^{AA'B'B}_{j,k}=\rho^{AA'}_j\otimes\rho^{B'B}_k$, $j,k=1,2$, is the initial joint state of the four ensembles, and $\chi_A^{\rho^{AA'}_j}$ and $\chi_A^{\rho^{B'B}_k}$ can be obtained from Eqs.~(\ref{CharFuncFirst}) and (\ref{CharFuncFirstN}). 

In order to perform a BSM on $A'$ and $B'$, in the DLCZ protocol, the states of these ensembles are transferred, using retrieval pulses \cite{DLCZ}, to optical modes on which a partial BSM is performed. Such a partial BSM has been schematically shown in Fig.~\ref{BSM} for two general input modes $X$ and $Y$. Here $\eta_c$ models the efficiency of atomic-to-photonic conversion, and $\eta_d$ is the quantum efficiency of photodetectors. The photodetectors in Fig.~\ref{BSM} are then assumed to have ideal unity quantum efficiencies. We assume that the coupling and quantum efficiencies are identical for both paths. The effect of the measurement module in Fig.~\ref{BSM} on the input-output characteristic functions is given by \cite{Razavi06}:
\begin{equation}\label{TransBSM}
\chi_A^{\rho^{X'Y'}}(\zeta_X,\zeta_Y) = \chi_A^{\rho^{XY}}(\sqrt{\eta_c} \zeta_X^-,\sqrt{\eta_c}\zeta_Y^+) B(\zeta_X,\zeta_Y),
\end{equation}
where $\rho^{XY}$ and $\rho^{ X'Y'}$ are, respectively, the state at the input ports $X$ and $Y$, and the state right before ideal detectors in Fig.~\ref{BSM}, and
\begin{eqnarray}
B(\zeta_X,\zeta_Y) &=& \exp[-(1-\eta_d)(|\zeta_X|^2+|\zeta_Y|^2)] \nonumber\\
&\times&\exp[-(1-\eta_c)(|\zeta_X^-|^2+|\zeta_Y^+|^2)]
\end{eqnarray}
with
\begin{equation}\label{zetaAzetaC}
\zeta_X^-=\sqrt{\frac{\eta_d}{2}}(\zeta_Y-\zeta_X)\ \ \mbox{and} \ \ 
\zeta_Y^+=\sqrt{\frac{\eta_d}{2}}(\zeta_Y+\zeta_X).
\end{equation}

In the repeater of Fig.~\ref{SetupSchematic}(b), optical modes retrieved from ensembles $A'$ and $B'$ go through a similar transformation to Eq.~(\ref{TransBSM}). Hence, for the resulting optical modes $A''$ and $B''$, we obtain
\begin{align}\label{TransBSMrepeater}\nonumber
\chi_A^{\rho^{ AA''B''B}_{j,k}}(\zeta_A,&\zeta_X,\zeta_Y,\zeta_{B}) = B(\zeta_X,\zeta_Y) \times \nonumber \\
&\chi_A^{\rho^{AA'B'B}_{j,k}}(\zeta_A, \sqrt{\eta_c}\zeta_X^-,\sqrt{\eta_c}\zeta_Y^+,\zeta_B),
\end{align}
where $\rho^{AA''B''B}_{j,k}$, $j,k=1,2$, is the joint density matrix of ensembles $A$ and $B$ and optical modes $A''$ and $B''$ in Fig.~\ref{SetupSchematic}(b). 

The measurement operators in the repeater scenario are
\begin{align}\label{Measurements}\nonumber
M_1=&|1\rangle_{A''A''}\langle 1|\otimes|0\rangle_{B''B''}\langle0|\\ 
M_2=&|0\rangle_{A''A''}\langle0|\otimes|1\rangle_{B''B''}\langle 1|
\end{align}
for the PNRD case, where $M_i$, $i=1,2$, corresponds to a single click on detector $D_i$, and $|i\rangle_K$, $i=0,1$ and $K=A'',B''$, represents a Fock state for the optical mode $K$. Similarly,
\begin{align}
M_1=&(I_{A''}-|0\rangle_{A''A''}\langle0|)\otimes|0\rangle_{B''B''}\langle0|\nonumber \\
M_2=&|0\rangle_{A''A''}\langle0| \otimes(I_{B''}-|0\rangle_{B''B''}\langle0|)
\end{align}
for the NRPD case, where $I_K$, $K=A'',B''$, represents the identity operator for mode $K$.

The final state of ensembles $A$ and $B$ will then be given by
\begin{equation}\label{RepeaterOutput}
\rho^{AB}_{i,j,k}=\frac{\mathrm{Tr}_{_{A''B''}}(M_i\rho^{AA''B''B}_{j,k})}{P_{i,j,k}^{\rm BSM}}, \quad\mbox{$i,j,k=1,2$},
\end{equation}
where 
\begin{align}\label{TransRhoChiRepeater}\nonumber
\rho^{AA''B''B}_{j,k}=&\int\frac{d^2\zeta_A}{\pi} \int\frac{d^2\zeta_{A''}}{\pi} \int\frac{d^2\zeta_{B''}}{\pi} \int\frac{d^2\zeta_B}{\pi} \nonumber \\  &\times\chi_A^{\rho^{AA''B''B}_{j,k}}(\zeta_A,\zeta_{A''},\zeta_{B''},\zeta_B) \nonumber\\
& \times D_N(S_A,-\zeta_A)D_N(a_{A''},-\zeta_{A''}) \nonumber\\
& \times D_N(a_{B''},-\zeta_{B''})D_N(S_B,-\zeta_B),
\end{align}
where $a_K$, $K=A'',B'' $, is the annihilation operator corresponding to the optical mode $K$, and 
\begin{equation}
\label{BSMherald}
P^{\rm BSM}_{{i,j,k}}=\mathrm{Tr}(M_i\rho^{A A''B''B}_{j,k}) \equiv P_M/2
\end{equation}
is the probability that only detector $D_i$, $i=1,2$, clicks in the BSM module. Because of the symmetry of our setup, this probability is independent of indexes $i$, $j$, and $k$, and it is half of the total BSM success probability $P_M$. Similar to Eq.~(\ref{FidelityGeneral}), we can define the fidelity for the final state as follows
\begin{equation}\label{Fidelityrepeater}
F^{(2)}_{AB}=\frac{1}{8}\sum_{i,j,k=1,2}\left._{AB}\langle\psi_{i+j+k}^{\rm Bell}|\rho^{AB}_{i,j,k}|\psi_{i+j+k}^{\rm Bell}\rangle_{AB}\right. .
\end{equation}
Given that the characteristic function in Eq.~(\ref{TransBSMrepeater}) has a Gaussian form, the above quantity can be turned into a Gaussian integral by plugging the following identities into Eqs.~(\ref{RepeaterOutput})--(\ref{Fidelityrepeater}). For any single-mode annihilation operator $a$ and complex variable $\zeta$, we have
\begin{eqnarray}
\label{(E13)}
&\langle 0 |D_N (a,\zeta )| 0 \rangle  = 1 \mbox{ , }   \langle 1 |D_N (a,\zeta )| 1 \rangle  = 1 - \left| \zeta  \right|^2, & \nonumber \\
&{\rm Tr}[ D_N (a,\zeta )] = \pi \delta (\zeta ),&
\end{eqnarray}
and, for any two ensembles $A$ and $B$, we have
\begin{align}
\label{(E17)} 
\langle \psi _j^{\rm Bell} | D_N (S_A ,\zeta _A ) D_N & (S_B ,\zeta _B )|\psi _j^{\rm Bell} \rangle  = \nonumber \\
& 1 - \frac{\displaystyle \left| \zeta _A  + ( - 1)^j \zeta _B  \right|^2 }{\displaystyle 2}.
\end{align}
As mentioned earlier, we use Maple to analytically simplify the resulting Gaussian integrals. The final result is, however, too long to be presented here.

\begin{figure}
\centering
{{\includegraphics[width =3in]{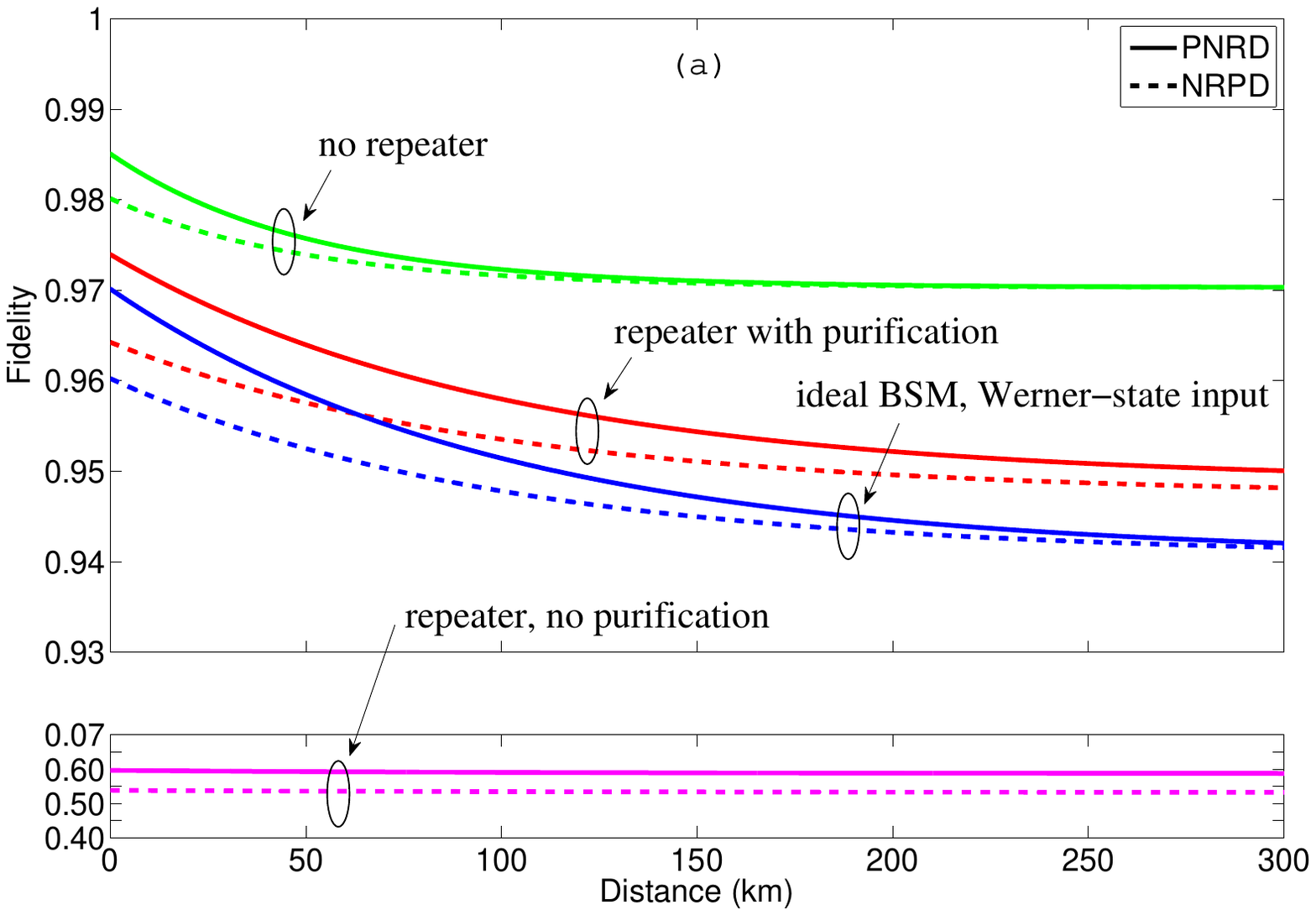}} \\ 
{\includegraphics[width=3in]{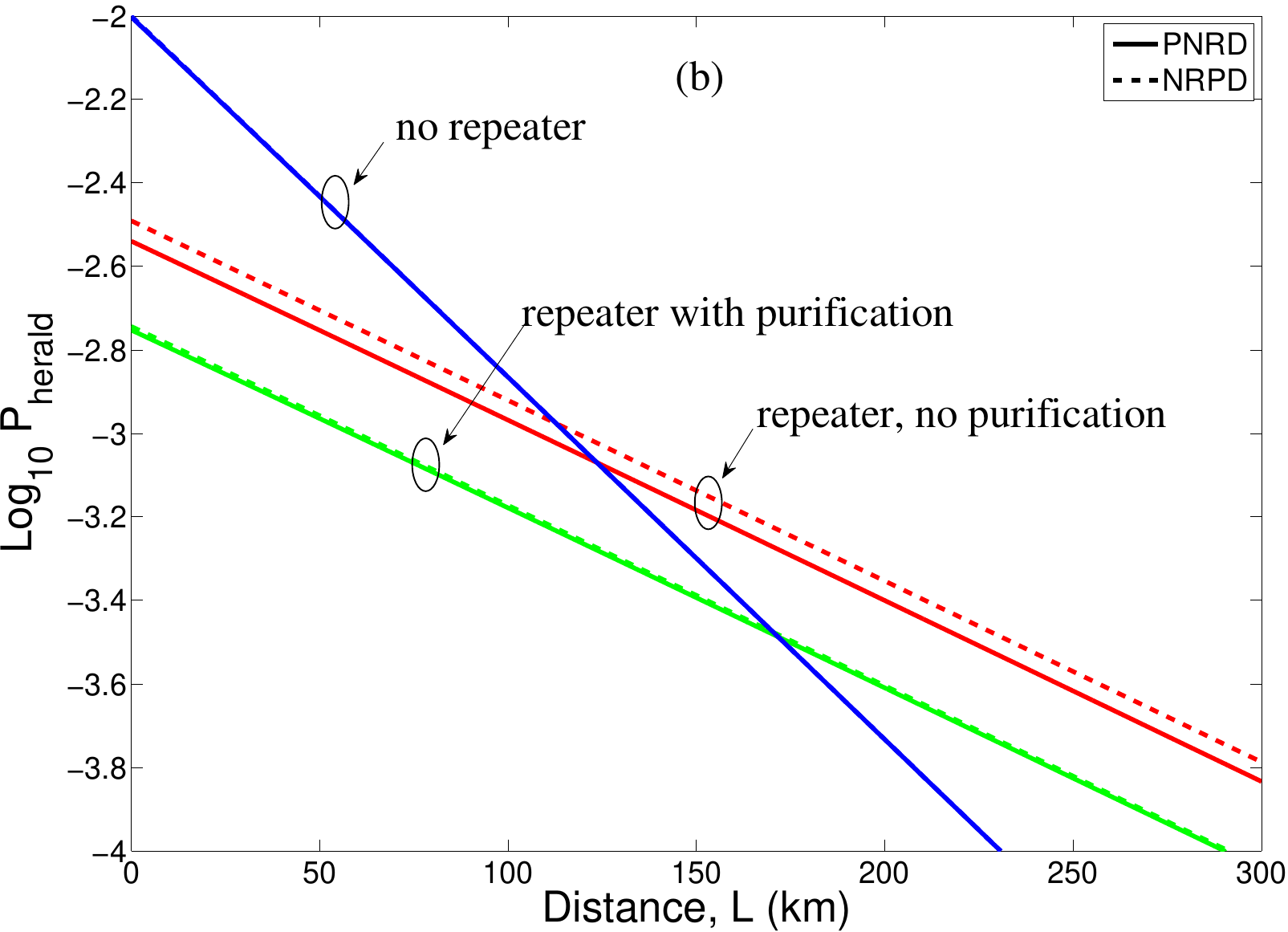}}
\caption{(Color online) (a) Fidelity versus distance, for ensembles $A$ and $B$ of Figs.~\ref{SetupSchematic}(a) and (b), in different scenarios using PNRDs (solid) and NRPDs (dashed). The ``no repeater'' curves represent the fidelity of the direct DLCZ entanglement distribution, $F^{(1)}_{AB}$, given by Eq.~(\ref{FidelityNorepeater}). The ``repeater, no purification'' curves are obtained from Eq.~(\ref{Fidelityrepeater}), and the ``repeater with purification'' curves are obtained from Eq.~(\ref{FidelityPurified}), when we exclude $|00\rangle_{AB}$ by post measurement. The final pair of curves represent the fidelity of entanglement assuming that the initial entangled states of $AA'$ and $BB'$ are Werner states with fidelity $F^{(1)}_{AA'} = F^{(1)}_{BB'}$, and that the employed BSM module is deterministic and error free. (b) The probability of heralding success for a DLCZ link that uses no repeater operation, $P_S(L)$, the one with one repeater node but no purification, $P_{S}(L/2) P_M$, and the one with one repeater node and self-purification, $P_{S}(L/2) P^{\rm purified}_{M}$. Again, plots with solid lines are for PNRDs and those with dashed lines are for the NRPDs. The parameter values, used in both plots, are $p_c=0.01$, $\eta_d = 0.5$, $\eta_c = 0.7$, and $L_{att} = 25\, {\rm km}$ corresponding to 0.17 dB/km loss in optical fibers. \label{Repeater}}}
\end{figure}

The fidelity obtained from Eq.~(\ref{Fidelityrepeater}) never exceeds $1/(2-\eta_m)$ for PNRDs and $1/(2-\eta_m/2)$ for NRPDs \cite{Razavi06}, and, therefore, at $\eta_m =0.35$, is substantially lower than that of direct entanglement distribution in Eq.~(\ref{FidelityNorepeater}); see Fig.~\ref{Repeater}(a). That is because the DLCZ repeater scheme is a conditional protocol. It can purify itself only after post-measurement processing. 

The main reason for the low fidelity of the DLCZ repeater is due to circumstances in which ensembles $A'$ and $B'$ each hold an excited atom, whereas ensembles $A$ and $B$ are in their vacuum states. In such a case, it is still possible that, in the presence of loss in the BSM module, we observe a single click on only one of the detectors, while the remote ensembles $A$ and $B$ are left in the state $|00\rangle_{AB}$. Ideally, such a vacuum state does not produce an error in the DLCZ QKD scheme because we need a minimum of two excitations in the four ensembles of Fig.~\ref{SetupSchematic}(c) to create one bit of the sifted key. It will be interesting then to look at the conditional fidelity when the final state is non-vacuum.

Suppose, we have performed a certain measurement by which we have learned that the joint state of $AB$ is non-vacuum. 
Such a purified density operator is then given by
\begin{align}\label{RhoPurified}
&\rho^{AB,{\rm purified}}_{i,j,k}=\nonumber \\
&\frac{(I-|00\rangle_{ABAB}\langle00|)\rho^{AB}_{i,j,k}(I-|00\rangle_{ABAB}\langle00|)} {\mathrm{Tr}[(I-|00\rangle_{ABAB}\langle00|)\rho^{AB}_{i,j,k}(I-|00\rangle_{ABAB}\langle00|)]}.
\end{align}
The fidelity of this new purified state will be given by
\begin{align}\label{FidelityPurified}
&F^{(2)}_{AB,{\rm purified}} =\nonumber \\ &\frac{1}{8}\sum_{i,j,k=1,2}\left._{AB}\langle\psi_{i+j+k}^{\rm Bell}|\rho^{AB,{\rm purified}}_{i,j,k}|\psi_{i+j+k}^{\rm Bell}\rangle_{AB}\right. ,
\end{align}
and the conditional probability that only one of the BSM detectors clicks, given that the final state of $A$ and $B$ is non-vacuum, is given by
\begin{equation}\label{PheraldPurified}
P^{\rm purified}_{M}= P_{M} -\sum_{i=1,2}\mathrm{Tr}(\rho^{AA''B''B}_{j,k}M_i|00\rangle_{ABAB}\langle00|).
\end{equation}

Figure~\ref{Repeater}(a) compares the fidelity of the direct DLCZ entanglement distribution given by Eq.~(\ref{FidelityNorepeater}) with three repeater scenarios. The first scenario is for the fidelity of the DLCZ repeater, given by Eq.~(\ref{Fidelityrepeater}), without considering its post-measurement self-purification property. As discussed before, this fidelity cannot be larger than $1/(2-\eta_m)$, and, for our employed parameters, it is around 0.6. However, if we assume that we can exclude the vacuum case later by post-measurement processing, the fidelity, given by Eq.~(\ref{FidelityPurified}), rises to about 0.95 and higher depending on the total distance; see the curves labeled ``repeater with purification'' in Fig.~\ref{Repeater}(a). The latter curves are still below that of the direct DLCZ link. This is to be expected because in the repeater protocol we start with non-ideal entangled states. In fact, even if we have an ideal error-free BSM module, and assuming that the initial fidelity $F_{AA'}^{(1)} = F_{BB'}^{(1)} = F$ corresponds to a Werner state, the fidelity of the entangled state of $A$ and $B$ after the ideal BSM goes down to $1/4 + (4/3)(F-1/4)^2$ \cite{Briegel98a}; see the curves labeled ``ideal BSM, Werner-state input'' in Fig.~\ref{Repeater}(a). The entangled state obtained by the DLCZ protocol is not a Werner state, and therefore, its fidelity drop cannot be modeled that way. The fidelity drop in the DLCZ case is slightly less than that of the Werner-state model as shown in Fig.~\ref{Repeater}(a). This improvement, however, is at the price of achieving a lower success rate due to employing a partial probabilistic BSM module in the DCLZ protocol.

Figure~\ref{Repeater}(b), shows the probability of heralding success in the two cases of with and without repeater. The heralding probability, in the repeater case, is defined as the product of $P_S(L_{AA'}) = P_S(L_{BB'}) = P_S(L/2)$, given by Eq.~(\ref{PjNorepeater}), for the initial entanglement distribution, and the BSM success probability, $P_M$ or $P^{\rm purified}_{M}$, given, respectively, by Eqs.~(\ref{BSMherald}) and (\ref{PheraldPurified}). It may seem, at the first glance, that the success rate in the quantum repeater case is proportional to $P_S(L_{AA'}) \times P_S(L_{BB'})$. It is not, however, the case because in a repeater setup, we do not perform the BSM before learning that the initial entanglement distribution has been successful on both links. On both $AA'$ and $BB'$ links, it takes on average about $1/P_S(L_{AA'}) = 1/P_S(L_{BB'})$ trials before they can perform the BSM on $A'B'$ ensembles. This descriptive argument can be made precise if we assume that at the end of each link, there is a bank of sufficiently large number of memories, on which this procedure is being successively attempted in parallel \cite{Razavi09a}. Under this assumption, it can be seen that, although, for short distances, the direct entanglement distribution has a better success rate, there is a crossing point at which the repeater protocol generates a higher number of entangled states. Notice that the quality of the entangled states generated by the repeater is lower than that of the DLCZ link. We deal with this issue and its implications on the rate later when we deal with a practical application, namely, QKD, in the following section. 

In Fig.~\ref{Repeater}, we have considered both cases of using PNRDs and NRPDs. As can be seen in Fig.~\ref{Repeater}(a), there is only a slight advantage in using resolving photodetectors for the purpose of entanglement distribution. For a fixed value of $p_c$, and at long distances, the fidelity in both cases approaches a similar constant value mostly determined by $p_c$. The heralding probability, however, is slightly higher in the NRPD case, and that is because, with NRPDs, two photons can masquerade themselves as a single photon. Such a scenario mostly occurs when the two middle ensembles hold two excited atoms altogether, and the remote ensembles are in their vacuum states. By excluding the vacuum state from the final state of the remote ensembles, our purified heralding probability is much less dependent on the type of employed detectors as shown in Fig.~\ref{Repeater}(b).

\section{Performance Analysis: Quantum Key Distribution}
\label{Sec:PA:QKD}
In this section, we obtain the secure key generation rate for the system shown in Fig.~\ref{SetupSchematic}(c) in two cases. First, when the initial entangled pairs are obtained from the direct DLCZ link of Fig.~\ref{SetupSchematic}(a), and, second, when the DLCZ repeater of Fig.~\ref{SetupSchematic}(b) is used. We use the same methodology as in the previous section by finding the relevant characteristic functions from which the final density matrices can be found. Any statistical moments of interest can then be written as Gaussian integrals. The final analytical results obtained by this method are, however, too long to fit in the paper and will be omitted.

The secure key generation rate is the product of three terms: the generation rate of entangled states to be employed in the QKD protocol, the probability that an acceptable click pattern occurs upon QKD measurements (denoted by $P_{\rm click}$ later in this section), and the ratio between the number of secure key bits and the sifted key bits. To obtain the first term, we use the results of \cite{Razavi09a} for the case of infinitely many memories, which states that for a quantum repeater with nesting level $n$, the generation rate of entangled states is given by $P_S(L/2^n)P_M^{(1)}P_M^{(2)} \cdots P_M^{(n)}/(2L/c)$, where $c$ is the speed of light in the channel and $P_M^{(i)}$, $i=1,\ldots, n$, is the BSM success probability at nesting level $i$. In our case, $P_S$ and $P_M$ were found in the previous section. To calculate the last term, we use the Shor-Preskill lower bound for the ratio between the number of secure key bits and the number of sifted key bits, in the limit of an infinitely long key, as given by \cite{preskill_shor}
\begin{equation}
\label{RBB84}
R_{\rm QKD}=1-2 H({\rm QBER}),
\end{equation}
where QBER is the quantum bit error rate, and ${H}(p)=-p\log_2p-(1-p)\log_2(1-p)$. The main assumption in deriving Eq.~(\ref{RBB84}) is that the QKD measurements are being performed on qubits. This assumption does not hold in our case because we are measuring infinite-dimensional optical modes, which cannot necessarily be modeled by qubits. It has recently been shown, however, that by using squashing techniques, the same key rate is achievable in our case as well \cite{Squash}. We then just need to obtain QBER and $P_{\rm click}$ to find the key generation rate as explained in the following.

\subsection{DLCZ QKD with no repeater}

Having entangled two pairs of ensembles, namely $A,B$ and $C,D$, via the DLCZ entanglement distribution protocol of Fig.~\ref{SetupSchematic}(a), the initial joint characteristic function of these four ensembles is as follows:
\begin{equation}\label{MultiplyChar}
\chi_A^{\rho^{ABCD}_{j}}(\zeta_A,\zeta_B,\zeta_C,\zeta_D)= \chi_A^{\rho^{AB}_j}(\zeta_A,\zeta_B)\times\chi_A^{\rho^{CD}_j}(\zeta_C,\zeta_D),
\end{equation}
where $\rho^{ABCD}_{j}=\rho^{AB}_j\otimes\rho^{CD}_j$, $j=1,2$, is the initial joint state of the four ensembles. In the DLCZ protocol, the atomic states are transferred to photonic states on which Alice and Bob perform their random QKD measurements by applying random phase shifts. Because they later discard the measurement results obtained from different phase shifts, we only consider the case where both Alice and Bob have chosen zero phase shifts. Under this assumption, the measurement modules on Alice and Bob's setup are identical to the one in Fig.~\ref{BSM}, and we can apply the transformation in Eq.~(\ref{TransBSM}) to obtain
\begin{align}
\label{TransBSMQKD}
\nonumber
\chi_A^{\rho^{A'B'C'D'}_{j}}&(\zeta_A,\zeta_B,\zeta_{C},\zeta_{D}) = B(\zeta_{A},\zeta_{C})B(\zeta_{B},\zeta_{D}) \times\\
&\chi_A^{\rho^{ABCD}_{j}}(\sqrt{\eta_c}\zeta_A^-, \sqrt{\eta_c}\zeta_B^-,\sqrt{\eta_c}\zeta_{C}^+,\sqrt{\eta_c}\zeta_{D}^+),
\end{align}
where $\zeta_A^-(\zeta_B^-)$ and $\zeta_C^+(\zeta_D^+)$ can be obtained from Eq.~(\ref{zetaAzetaC}) by replacing $X$ with $A(B)$ and $Y$ with $C(D)$. The density operator $\rho^{A'B'C'D'}_{j}$ represents the state of the optical modes right before their being detected by the ideal single photon detectors in Fig.~1(c), and is given by
\begin{align}\label{TransRhoChi4}\nonumber
\rho^{A'B'C'D'}_{j}=&\int\frac{d^2\zeta_A}{\pi} \int\frac{d^2\zeta_B}{\pi} \int\frac{d^2\zeta_C}{\pi} \int\frac{d^2\zeta_D}{\pi}\\
 \nonumber &\times\chi_A^{\rho^{A'B'C'D'}_{j}}(\zeta_A,\zeta_B,\zeta_C,\zeta_D)\\
 & \times D_N(a_{A'},-\zeta_A) D_N(a_{B'},-\zeta_B) \nonumber\\
&\times D_N(a_{C'},-\zeta_C)D_N(a_{D'},-\zeta_D),
\end{align}
where $a_K$, $K=A',B',C',D'$, is the annihilation operator corresponding to the optical mode $K$.

We use measurement operators to model the relevant QKD measurements. The most general measurement operator for the PNRD case is given by
\begin{equation}\label{MeasurementsQKD}
M_{abcd}=|a\rangle_{A'A'}\langle a|\otimes|b\rangle_{B'B'}\langle b|\otimes|c\rangle_{C'C'}\langle c|\otimes|d\rangle_{D'D'}\langle d|,
\end{equation}
for $a,b,c,d=0,1$, where $|k\rangle_K$ represents a Fock State for the optical mode $K=A',B',C',D'$. In the case of NRPDs, we only need to replace $|1\rangle_{KK}\langle 1|$ with $(I_K-|0\rangle_{KK}\langle0|)$, for $K=A',B',C',D'$, where $I_K$ is the identity operator for system $K$.

Let us consider the case of $j=1$ and denote the probability that $M_{abcd}$ occurs by
\begin{equation}
P_{abcd}=\mathrm{Tr}(\rho^{A'B'C'D'}_{1}M_{abcd}).
\end{equation}
Such probabilities can be calculated using the Gaussian integral techniques along with Eq.~(\ref{(E13)}). Because of the symmetry assumption, the case of $j=2$ will provide us with the same result in the end. The QBER is then given by
\begin{equation}\label{QBER}
\mathrm{QBER}=\frac{P_{\rm error}}{P_{\rm click}},
\end{equation}
where 
\begin{eqnarray}
\label{Pclick}
&\!&{P_{\rm click}=}\nonumber \\
&\!& \left\{ \begin{array}{cr} P_{1001}+P_{0110} + P_{1010}+ P_{0101}, & \mbox{PNRD},  \\
1-P_{1000} - P_{0100}- P_{0010} - P_{0001} - P_{0000}, & \mbox{NRPD},
\end{array} \right. \nonumber \\
\end{eqnarray}
is the probability of getting at least one click on each side, and 
\begin{equation}\label{Perror}
P_{\rm error}= \left\{ \begin{array}{cr} P_{1001}+P_{0110},& \quad\mbox{PNRD}, \\
P_{1001}+P_{0110}+\frac{1}{2}P_{0111}+\frac{1}{2}P_{1011} & \\
+\frac{1}{2}P_{1101}+\frac{1}{2}P_{1110}+\frac{1}{2}P_{1111},&\quad\mbox{NRPD},
\end{array} \right.
\end{equation}
is the probability of making an error, i.e., Alice and Bob assign different bits to their sifted keys. In the PNRD case, we only count cases where exactly one photon has been detected \cite{TobiNJP}. So, a bit error occurs whenever there is a mismatch between detectors that have clicked. In the NRPD case, we have to only consider the double-click cases, where we assign a random bit to the sifted key. The terms that start with a $1/2$ factor account for the probability of error in the double-click cases. 

Assuming that the DLCZ protocols for entanglement distribution and QKD are being successively applied, with period $L/c$, to a large number of memories, the number of secure key bits generated per second per logical memory used in the system is lower bounded by \cite{Razavi09a}
\begin{equation}
\label{R1}
R_1=\frac{1-2 {H}(\mathrm{QBER})}{2L/c}\cdot P_S(L)\cdot P_{\rm click}/2.
\end{equation}
In the above equation, $P_S(L)/(2L/c)$ is the generation rate of entangled pairs per logical memory, $P_{\rm click}/2$  represents our likelihood of creating a sifted key bit by using two entangled pairs, and $1-2 {H}(\mathrm{QBER})$ represents the number of secure key bits created out of each sifted key bit. Here, we have neglected the cases where Alice and Bob choose different phase shift values. It has been shown that in order to detect an eavesdropper, it suffices for Alice and Bob to use different phase shifts with a probability that can approach zero \cite{QKDbiased}. How, in practice, this probability is chosen depends on the employed privacy amplification and reconciliation protocols. 

\subsection{DLCZ QKD with One Repeater Node}

Suppose the entangled states of $AB$ and $CD$, in Fig.~\ref{SetupSchematic}(c), are provided by the DLCZ repeater scheme of Fig.~\ref{SetupSchematic}(b). In order to find the rate, similar to the previous section, we first need to find the initial characteristic function for the composite system of ensembles $A$, $B$, $C$, and $D$. That will be given by 
\begin{equation}\label{MultiplyCharRep}
\chi_A^{\rho^{ABCD}_{i,j,k}}(\zeta_A,\zeta_B,\zeta_C,\zeta_D)= \chi_A^{\rho^{AB}_{i,j,k}}(\zeta_A,\zeta_B)\times\chi_A^{\rho^{CD}_{i,j,k}}(\zeta_C,\zeta_D),
\end{equation}
where $\rho^{ABCD}_{i,j,k} = \rho^{AB}_{i,j,k} \otimes \rho^{CD}_{i,j,k}$, $i,j,k=1,2$, is the initial joint state of the four ensembles with $\rho^{AB}_{i,j,k}$ given by Eq.~(\ref{RepeaterOutput}). Here, for simplicity, we assumed that the original entangled states are identical for both $AB$ and $CD$ systems. Other cases can be converted to this case by applying a local unitary operation. The next step is to calculate $\chi_A^{\rho^{AB}_{i,j,k}}(\zeta_A,\zeta_B)$, which will be dealt with in Appendix. The rest of the rate analysis then follows from Eqs.~(\ref{TransBSMQKD})--(\ref{Perror}) with obvious replacements and will be omitted. Again, under the assumption of large number of memories and parallel successive entangling attempts, the number of secure key bits, in the limit of long key, generated per second per logical memory used in the system is lower bounded by
\begin{equation}
\label{R2}
R_2=\frac{1-2 {H}(\mathrm{QBER})}{2L/c}\cdot P_S(L/2) \cdot P_M \cdot P_{\rm click}/2.
\end{equation}

\subsection{Numerical Comparison}

\begin{figure}
\centering
{{\includegraphics[width=2.5in]{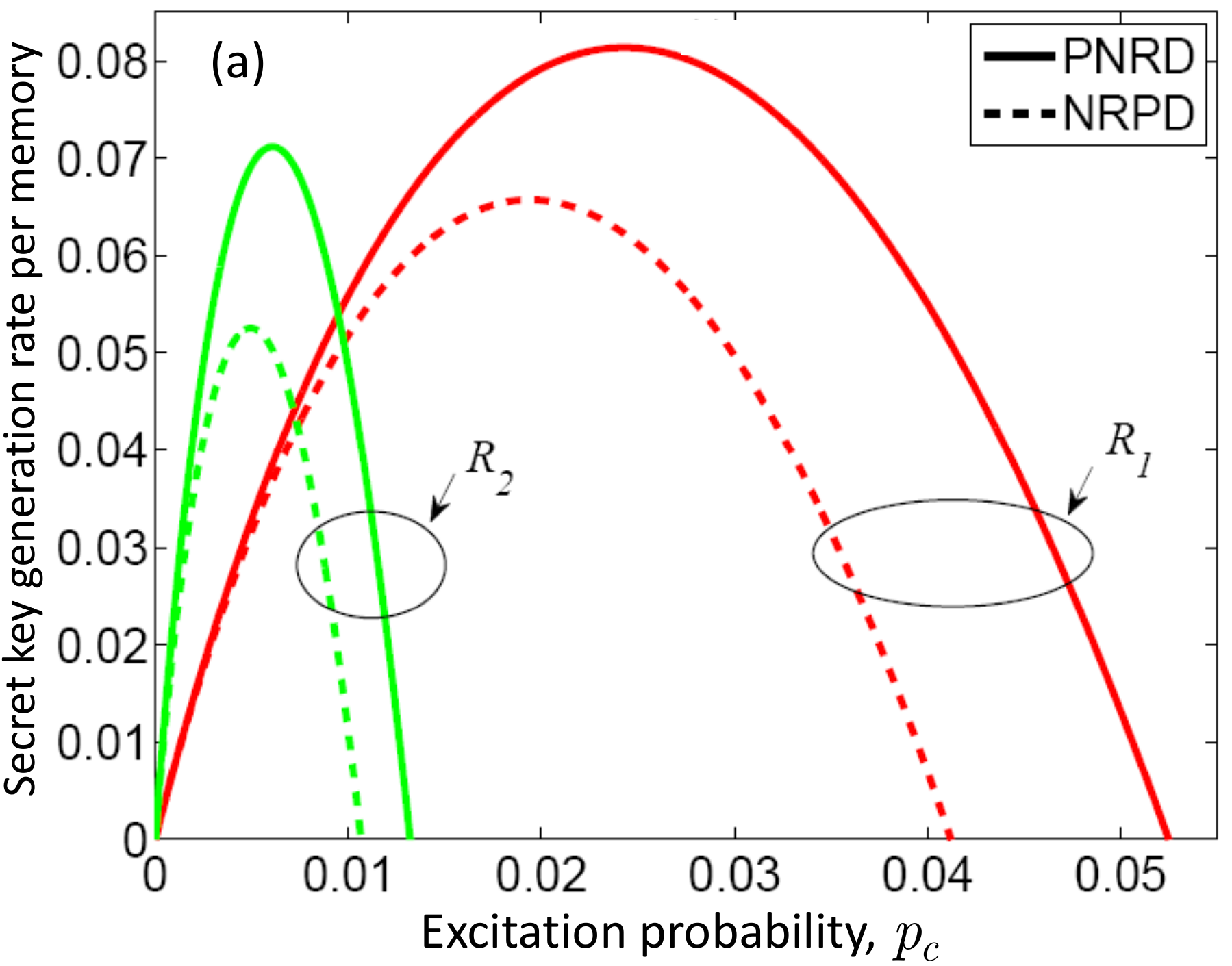}}\\ 
{\includegraphics[width =2.5in]{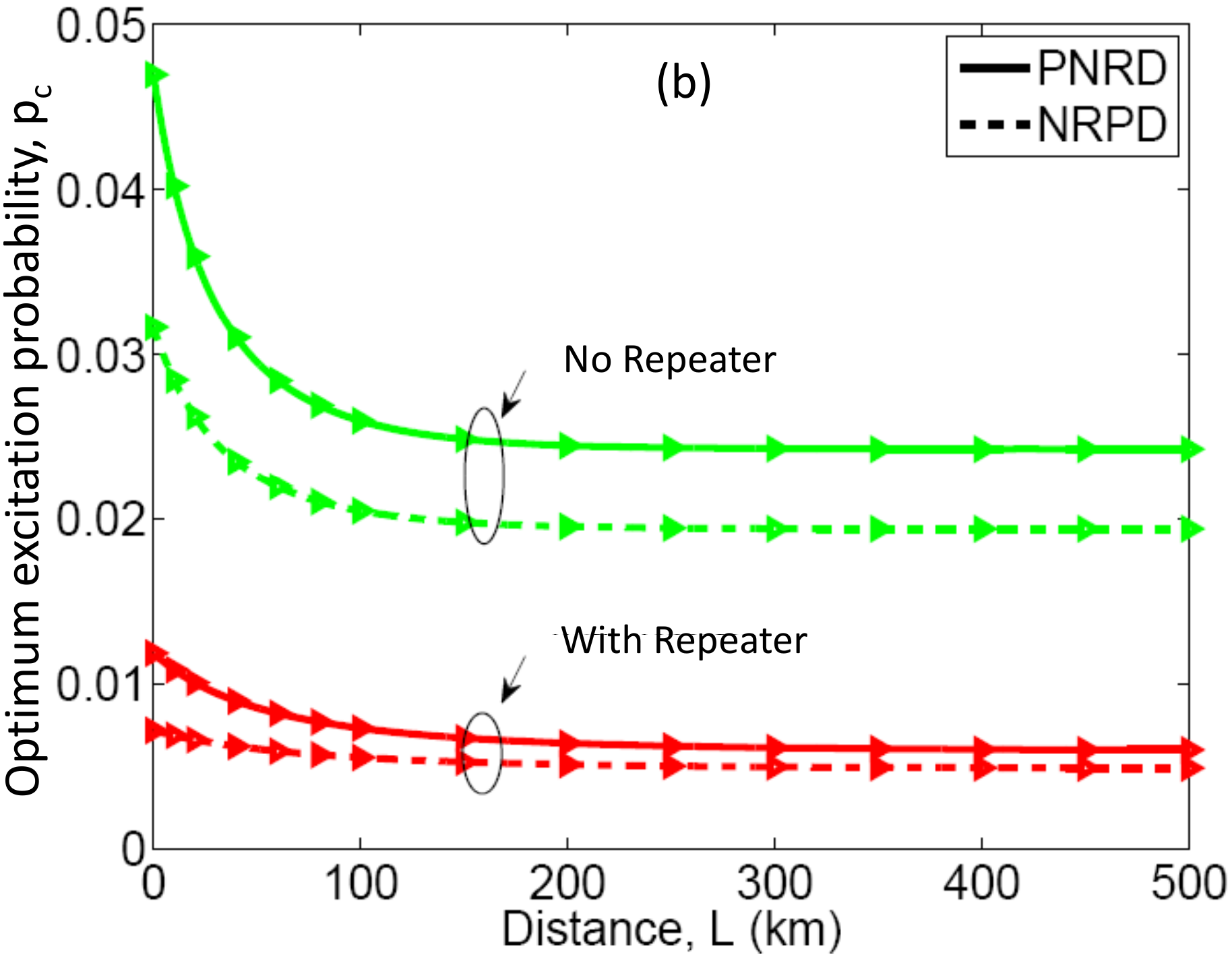}}
\caption{(Color online) (a) The number of secure key bits generated per second per logical memory for a DLCZ QKD system, with ($R_2$) and without ($R_1$) a repeater node, versus $p_c$ at $L=350$ km. There exists an optimum value of $p_c$ at which the QKD rate peaks. (b) Optimum value of $p_c$ versus distance. For large distances, roughly greater than 100 km, this optimum value approaches a constant value in each configuration. In (a) and (b), solid lines are for PNRDs and dashed lines are for NRPDs. The parameter values, used in both plots, are $\eta_d = 0.5$, $\eta_c = 0.7$, and $L_{att} = 25$~km and $c=2 \times 10^8$ m/s for the optical fiber channel. \label{QKDrate}}}
\end{figure}

In this section we compare the normalized rate given by Eq.~(\ref{R1}) for the no-repeater QKD link with that of Eq.~(\ref{R2}) for the single-hop repeater configuration. We find the dependence of $R_1$ and $R_2$ on various system parameters such as the excitation probability $p_c$, the total distance $L$, and the measurement efficiency $\eta_m$.

Figure~\ref{QKDrate}(a) shows $R_1$ and $R_2$ as functions of the excitation probability, $p_c$, for a 350-km-long optical fiber channel with 0.17 dB/km loss. It can be seen that there exist optimum values of $p_c$ at which the QKD rates peak. That is because, whereas a higher value of $p_c$ increases the heralding rate of success, it also creates more multiple-excitation errors, which, in turn, reduces the rate. The optimum value of $p_c$ for the repeater setup is lower than that of the no-repeater case. It is also lower for non-resolving detectors than the resolving ones. That is because with the repeater system, or, with the non-resolving detectors, we create more errors in the entanglement swapping/distribution steps, hence we are better off to start with a lower value of $p_c$ to allow for a higher margin of error by the end of the QKD procedure.

\begin{figure}
\centering
{{\includegraphics[width =2.5in]{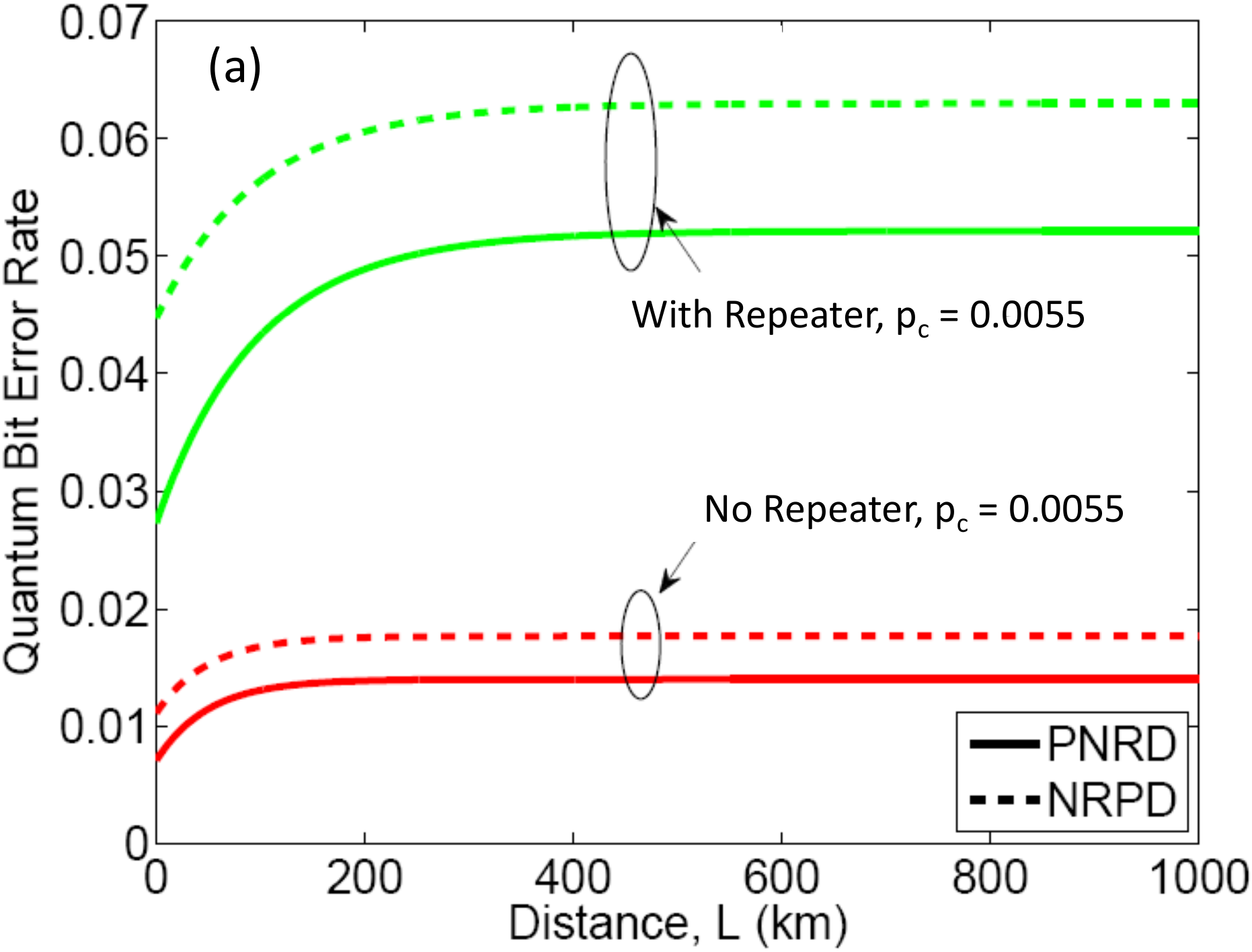}}\\ 
{\includegraphics[width=2.5in]{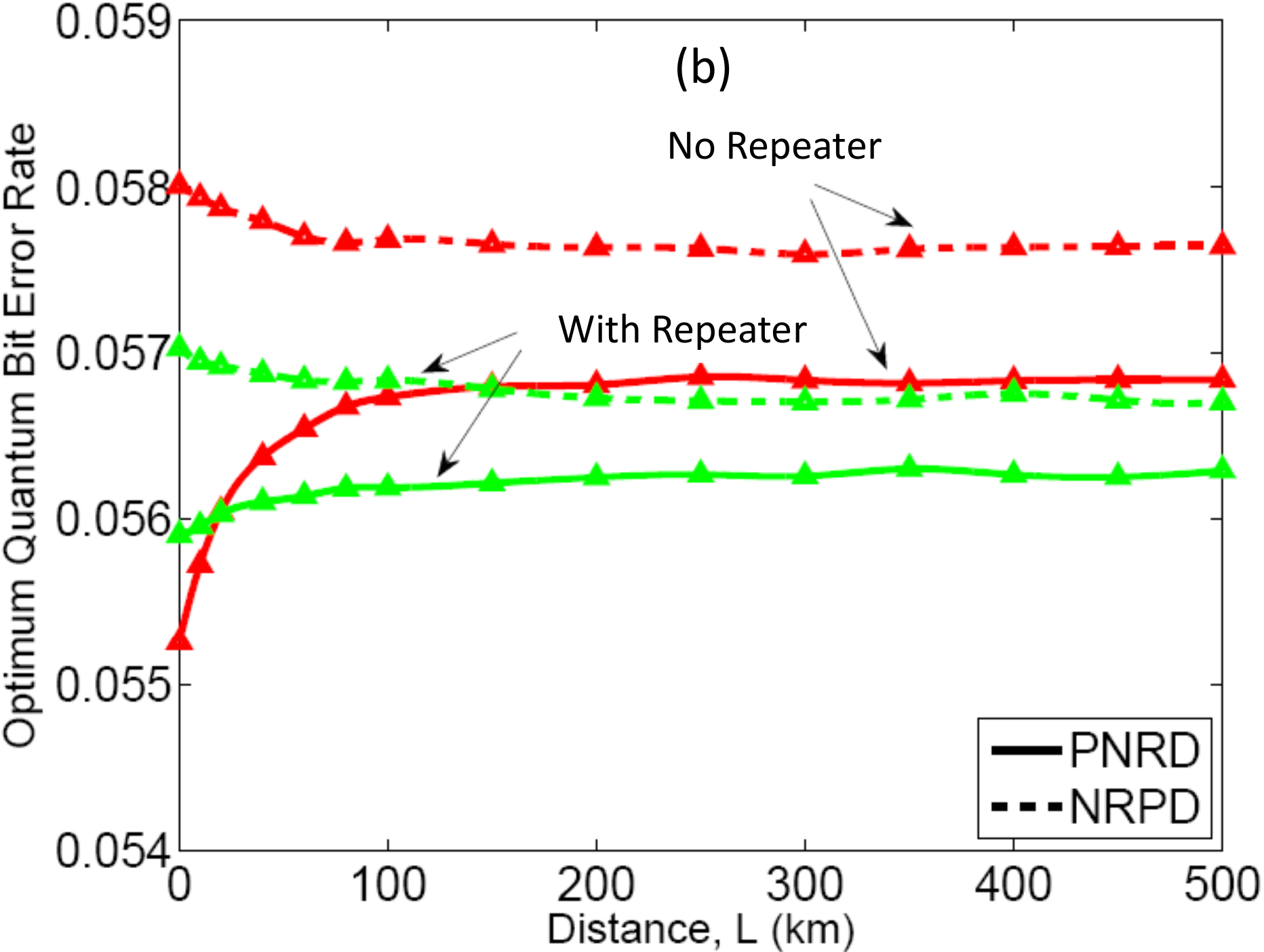}}
\caption{(Color online) (a) QBER versus distance for the DLCZ QKD protocol, with and without repeater, at $p_c=0.0055$. At an identical value for $p_c$, the repeater systems undergo higher error rates. This is also the case for systems that use NRPDs (dashed lines) rather than PNRDs (solid lines). (b) QBER versus distance for the DLCZ QKD protocol, with and without repeater, at the optimum values of $p_c$. QBER is about the same for all configurations. The parameter values used in both plots are $\eta_d = 0.5$, $\eta_c = 0.7$, and $L_{att} = 25\, {\rm km}$. \label{QBERFIG}}}
\end{figure}

The optimum values of $p_c$ are functions of distance as well. Figure~\ref{QKDrate}(b) shows such dependence for resolving and non-resolving detectors with and without a repeater node. It can be seen that, whereas, for short distances, a higher value of $p_c$ is desired, the optimum value of $p_c$ approaches a certain value in the limit of long distances. For our parameter setting, with nominal values of $\eta_d = 0.5$ and $\eta_c = 0.7$, the optimum values are roughly constant for $L>100$ km. These long-distance-limit optimum values are $p_c = 0.0243$ and $p_c = 0.0194$ for the no-repeater case using, respectively, PNRDs and NRPDs, and $p_c = 0.0060$ (PNRD) and $p_c = 0.0049$ (NRPD) for the configuration with one repeater node. 

The asymptotic behavior of the optimal value of $p_c$ is a result of the compromise between two competing terms in Eqs.~(\ref{R1}) and ({R2}). The first term is $1-2H({\rm QBER})$, which is a decreasing function of $p_c$ as higher values of $p_c$ increase the chance of multiple excitations. The second important term is the heralding probability $P_S$, which is an increasing function of $p_c$. The optimum value of $p_c$ is where these two competing factors balance each other.

Figure~\ref{QBERFIG}(a) shows QBER versus distance at $p_c = 0.0055$. As expected, at an identical value of $p_c$, the error rate for repeater systems is higher than the no-repeater ones. NRPD-based systems are also more prone to creating errors than the systems that use PNRDs. At short distances, the QBER is less than its asymptotic limit at long distances, which is because the fidelity decreases with distance; see Fig.~\ref{Repeater}(a). This lower error will enable us to increase $p_c$ to achieve a higher rate at lower distances. That explains why the optimum value of $p_c$ in Fig.~\ref{QKDrate}(b) is a decreasing function of distance. By increasing $p_c$ to its optimum value at each distance we expect to get a flat curve for QBER, as shown in Fig.~\ref{QBERFIG}(b). Interestingly, QBER is about the same for all configurations if we use the corresponding optimum value of $p_c$ for each setup at each distance. It is higher, however, for NRPD-based setups as compared to PNRD-based ones. 

\begin{figure}
\centering
{{\includegraphics[width =2.5in]{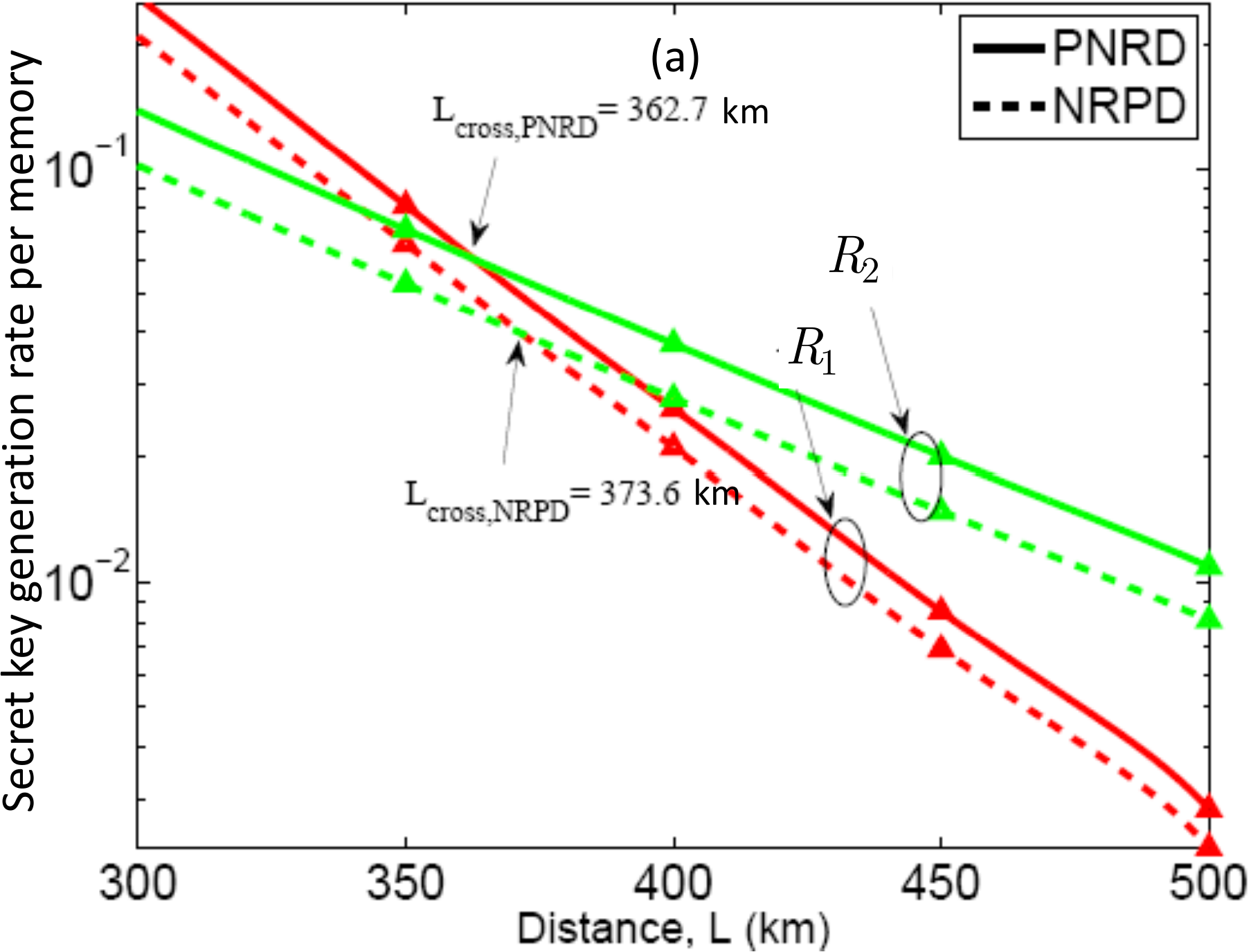}}\\ 
{\includegraphics[width=2.5in]{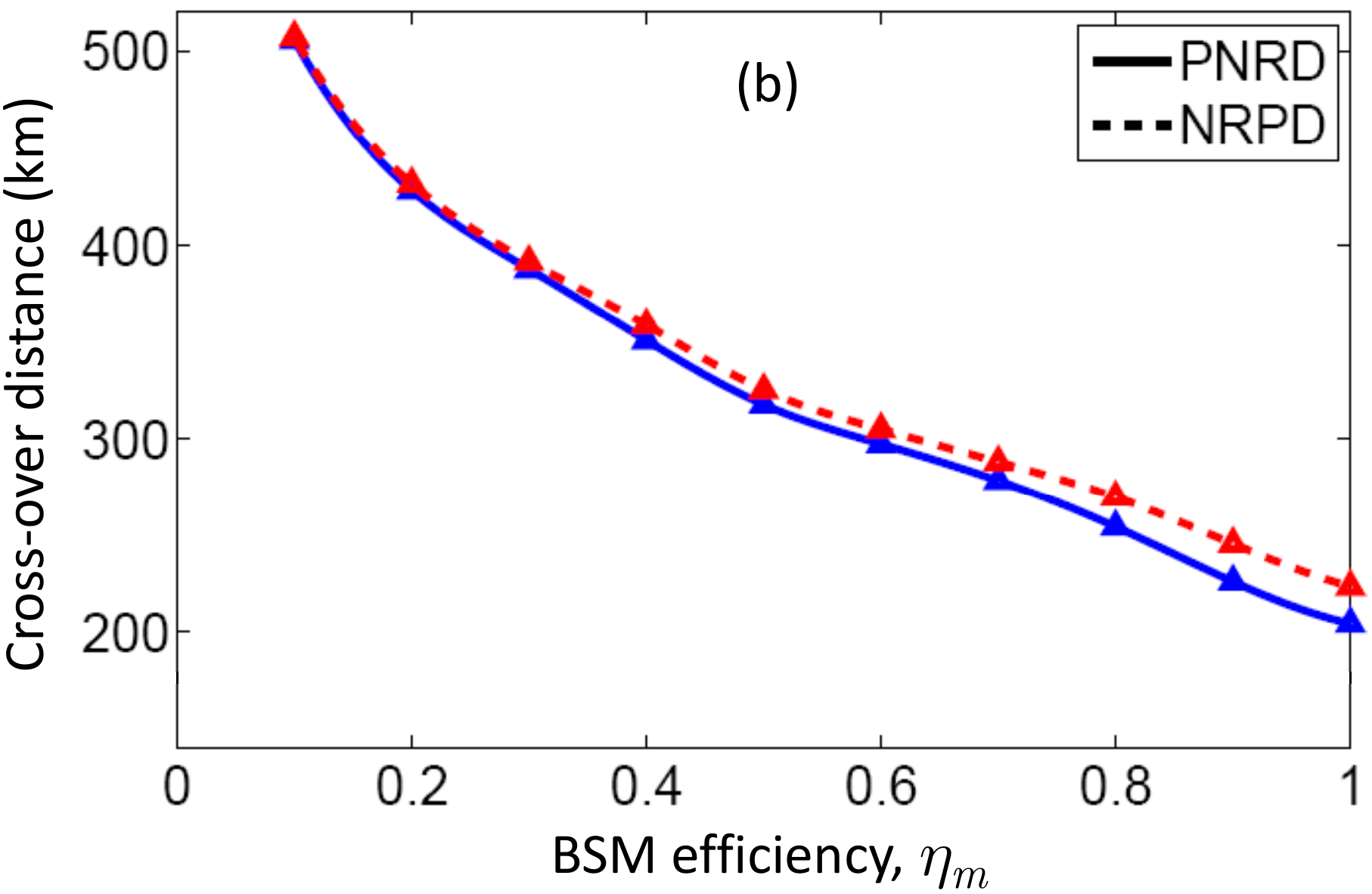}}
\caption{(Color online) (a) The number of secure key bits generated per second per logical memory used in the DLCZ QKD protocol, with and without a repeater node, versus distance at the optimum values of $p_c$. The cross-over distance, $L_{\rm cross}$, beyond which the repeater system outperforms the no-repeater system, has been shown for both NRPD-based (dashed) and PNRD-based (solid) systems at $\eta_d=0.5$ and $\eta_c=0.7$. (b) The cross-over distance as a function of the measurement efficiency, $\eta_m$, for the repeater BSM module. For the initial entanglement distribution, $\eta_d=0.5$. With higher values of $\eta_m$, the BSM success probability improves, and therefore the optimum distancing between repeater nodes will decrease. In both plots $L_{att} = 25$ km and $c = 2 \times 10^8$ m/s. \label{QKDcross}}}
\end{figure}

Finally, Fig.~\ref{QKDcross}(a) shows $R_1$ and $R_2$ versus distance. Each point on the graph is calculated at the corresponding optimum value of $p_c$. Because the optimum is almost identical, at long distances, for both repeater and non-repeater setups, the former eventually outperforms the latter, because of its higher efficiency at long distances. For the parameters used in our calculations, the cross-over distance, i.e., the distance beyond which the repeater setup outperforms the direct DLCZ link, is around 350 km. Note that because we are looking at {\em normalized} rates, the cost of extra memories used in a repeater setup is also included in our comparison. It is interesting to note that this cross-over distance for the QKD case is almost twice as large as what we found for the heralding probabilities in Fig.~\ref{Repeater}(b). While it is an unexpected result, it shows the importance of including the actual application in mind when such cross-over distances are calculated.

The cross-over distance also determines the optimum distancing between quantum repeater nodes. The latter depends on the BSM success probability $P_M$, which is a function of our measurement efficiency $\eta_m$. Figure~\ref{QKDcross}(b) shows the cross-over distance versus $\eta_m$ for $R_1$ and $R_2$ calculated at the optimum values of $p_c$. For highly efficient BSM modules, this distance will drop to about 200 km as shown in Fig.~\ref{QKDcross}(b). That is about 100 km between each repeater station. This is a characteristic of quantum repeater systems with probabilistic components, which tend to have only a few intermediate stations as compared to systems that rely on deterministic BSMs. The key generation rate is almost linearly growing with the number of logical memories employed \cite{SPIE}. According to Fig.~\ref{QKDcross}(a), to create 1 kbps of secure key, at $\eta_m = 0.35$, over 400 km, we need a total of about 30,000 logical memories for the DLCZ system. That could translate into 300--3000 physical memories in the system.

\section{Conclusions}
\label{Sec:Con}
Quantum communication systems that rely on probabilistic schemes for entanglement distribution and swapping provide us with a practical route towards long-distance quantum cryptography. In the scheme proposed by Duan, Lukin, Cirac, and Zoller, the interaction of light with the memory is enhanced via a collective excitation process. Multiple excitations are then the fundamental sources of error in such systems. In this paper, we considered the DLCZ proposal for quantum repeaters and quantum key distribution and calculated the generation rate of secure key bits per logical memory used in the system. The number of employed memories is a measure of cost in quantum repeaters. It turned out that in order to obtain the maximum rate for such systems, the collective excitation probability must be tuned to an optimal value. Such an optimal value was not sensitive to the total distance in the limit of long distances. We also compared the key generation rate for a DLCZ system that used a quantum repeater node for entanglement distribution with a direct no-repeater DLCZ system, and found the cross-over distance beyond which the quantum repeater outperformed the no-repeater system. The cross-over distance was about hundreds of kilometers, and that would vary depending on the efficiency of system components. That implied that probabilistic quantum repeaters might only need a few intermediate stations for entanglement swapping. That could reduce the cost of implementing such systems and increase their compatibility with current optical infrastructure. Another cost-saving observation was the fact that using photon-number resolving detectors only slightly improved system performance, and, any practical system could rely on non-resolving photodetectors at no appreciable loss in efficiency. To obtain a reasonable key rate, however, a large number of logical memories was required in each station.

\section*{ACKNOWLEDGMENTS} 
The second author would like to thank N. L\"utkenhaus and T. Moroder for fruitful discussions. This work was supported by QuantumWorks, OCE, and NSERC Discovery Grant.

\appendix*
\section{}
In this Appendix, we obtain the characteristic function for the output state of the DLCZ quantum repeater, $\rho^{AB}_{i,j,k}$, as given in Eq.~(\ref{RepeaterOutput}). Using the general definition of Eq.~(\ref{CharFuncDef}) and the identity ${\rm Tr}(D_A(a,\zeta)D_N(a,\zeta')) = \pi \delta(\zeta + \zeta')$, after some algebraic simplification, we obtain for PNRDs
\begin{align}\label{CharFuncSecond} \nonumber 
&\chi_A^{\rho_{i,j,k}^{AB}}(\zeta_A,\zeta_B) = {\rm Tr}( \rho^{AB}_{i,j,k} D_A(S_A,\zeta_A)D_A(S_B,\zeta_B)) \\ \nonumber
&=\exp[-\alpha(|\zeta_A|^2+|\zeta_B|^2)][1+c_1|\zeta_A|^2|\zeta_B|^2 \\
&+c_2(|\zeta_A|^2+|\zeta_B|^2)+c_3|\zeta_A+(-1)^{i+j+k}\zeta_B|^2],
\end{align}
where
\begin{align}\nonumber
c_1 = &[p_c({\eta_s}-1)(-1-\eta_mp_c+p_c-\eta_sp_c+\eta_m\eta_sp_c)^2]/\\ \nonumber
&[(2\eta_s\eta_mp_c-2\eta_mp_c+\eta_m-2+2p_c-2\eta_sp_c)\\ \nonumber
&(1+\eta_sp_c-p_c)^2(2\eta_s^2\eta_mp_c^2-4\eta_s\eta_mp_c^2+2\eta_mp_c^2+\\ \nonumber &\eta_s\eta_mp_c-\eta_mp_c-2\eta_s^2p_c^2+4\eta_sp_c^2-\eta_sp_c-2p_c^2\\ \nonumber
&+p_c+1)]\\ \nonumber
c_2=&[-p_c(\eta_s-1)(-1-\eta_mp_c+p_c-\eta_sp_c+\eta_m\eta_sp_c)]/\\ \nonumber
&[(1+\eta_sp_c-p_c)(2\eta_s^2\eta_mp_c^2-4\eta_s\eta_mp_c^2+2\eta_mp_c^2+\\ \nonumber &\eta_s\eta_mp_c-\eta_mp_c-2\eta_s^2p_c^2+4\eta_sp_c^2-\eta_sp_c-2p_c^2\\ \nonumber
&+p_c+1)]\\ \nonumber
c_3=&[-(-1-\eta_mp_c+p_c-\eta_sp_c+\eta_m\eta_sp_c)]/\\ \nonumber
&[2(2\eta_s\eta_mp_c-2\eta_mp_c+\eta_m-2+2p_c-2\eta_sp_c)\\ \nonumber
&(2\eta_s^2\eta_mp_c^2-4\eta_s\eta_mp_c^2+2\eta_mp_c^2+\eta_s\eta_mp_c\\ 
&-\eta_mp_c-2\eta_s^2p_c^2+4\eta_sp_c^2-\eta_sp_c-2p_c^2+p_c+1)],
\end{align}
and $\eta_s = \eta_s^{AA'}$, and for the NRPDs
\begin{align}\nonumber
&\chi_A^{\rho_{i,j,k}^{AB}}(\zeta_A,\zeta_B) = {\rm Tr}( \rho^{AB}_{i,j,k} D_A(S_A,\zeta_A)D_A(S_B,\zeta_B)) \\ \nonumber
&=c_1\exp(p_1(|\zeta_A|^2+|\zeta_B|^2) +p_2(|\zeta_A+(-1)^{i+j+k}\zeta_B|^2))\\ \nonumber
&+c_2\exp(p_1(|\zeta_A|^2+|\zeta_B|^2)) +c_3\exp(p_3|\zeta_A|^2+p_4|\zeta_B|^2)\\ \nonumber
&+c_4\exp(p_1|\zeta_A|^2+p_4|\zeta_B|^2) +c_3\exp(p_4|\zeta_A|^2+p_3|\zeta_B|^2)\\
&+c_4\exp(p_4|\zeta_A|^2+p_1|\zeta_B|^2)+c_5\exp(p_4(|\zeta_A|^2+|\zeta_B|^2)),
\end{align}
where
\begin{align}\nonumber
p_1=&(-\eta_sp_c+\eta_s\eta_mp_c-2\eta_mp_c+2p_c-2)/ \\ \nonumber &(2\eta_mp_c-\eta_s\eta_mp_c+2-4p_c+2\eta_sp_c-2\eta_sp_c^2+2p_c^2 \\ \nonumber &+2\eta_s\eta_mp_c^2-2\eta_mp_c^2) \\ \nonumber
p_2 = &(\eta_s^2\eta_mp_c^2)/[4(-1+p_c)(1+\eta_sp_c-p_c)(2\eta_mp_c-\eta_s\eta_mp_c\\ \nonumber
&+2\eta_s\eta_mp_c^2-2\eta_mp_c^2+2-4p_c+2\eta_sp_c-2\eta_sp_c^2+p_c^2)]\\ \nonumber
p_3=&(6p_c^2\eta_s+3p_c\eta_s\eta_m-6\eta_s\eta_mp_c^2-4-2\eta_s^2p_c^2-6\eta_sp_c\\ \nonumber &+4\eta_mp_c^2-4\eta_mp_c+2\eta_s^2\eta_mp_c^2-4p_c^2+8p_c)/  \\ \nonumber &((1+\eta_sp_c-p_c)(-3\eta_s\eta_mp_c+4\eta_mp_c+4-8p_c \\ \nonumber &+4\eta_sp_c-4\eta_sp_c^2+4p_c^2+4\eta_s\eta_mp_c^2-4\eta_mp_c^2)) \\
p_4=&\frac{-1}{1+\eta_sp_c-p_c}
\end{align}
and
\begin{align}
c_1=&(-2(-1+p_c)(1+\eta_sp_c-p_c)^3)/(\eta_s^2p_c^2(2\eta_mp_c \nonumber \\ \nonumber
&-\eta_s\eta_mp_c+2-4p_c+2\eta_sp_c-2\eta_sp_c^2+2p_c^2 \\ \nonumber &+2\eta_s\eta_mp_c^2-2\eta_mp_c^2)) \\ \nonumber
c_2=&(-4(-1+p_c)(1+\eta_sp_c-p_c)^3(-1-\eta_sp_c+2p_c \\ \nonumber &+\eta_sp_c^2-p_c^2))/(\eta_s^2p_c^2(2\eta_mp_c-\eta_s\eta_mp_c+2-4p_c \\ \nonumber
&+2\eta_sp_c-2\eta_sp_c^2+2p_c^2+2\eta_s\eta_mp_c^2-2\eta_mp_c^2)) \\ \nonumber
c_3=&(-4(1+\eta_sp_c-p_c)^2(-1+p_c)^2)/(\eta_s^2p_c^2\\ \nonumber &(-3\eta_s\eta_mp_c+4\eta_mp_c+4-8p_c+4\eta_sp_c-4\eta_sp_c^2 \\ \nonumber &+4p_c^2+4\eta_s\eta_mp_c^2-4\eta_mp_c^2)) \\ \nonumber
&-\eta_s\eta_mp_c+2-4p_c+2\eta_sp_c-2\eta_sp_c^2+2p_c^2 \\ \nonumber &+2\eta_s\eta_mp_c^2-2\eta_mp_c^2)) \\ \nonumber
c_4=&(-2(1+\eta_sp_c-p_c)^3(-1+p_c)^2)/((2\eta_mp_c-\eta_s\eta_mp_c \\ \nonumber &+2-4p_c+2\eta_sp_c-2\eta_sp_c^2+2p_c^2+2\eta_s\eta_mp_c^2 \\ \nonumber &-2\eta_mp_c^2)(-1-\eta_sp_c+p_c+\eta_s\eta_mp_c\eta_mp_c)\eta_s^2p_c^2) \\ \nonumber
c_5=&(-(1+\eta_sp_c-p_c)(-1+p_c)^2\eta_m(\eta_s-1))/\\ 
&(\eta_s^2p_c(-1-\eta_sp_c+p_c+\eta_s\eta_mp_c-\eta_mp_c)^2).
\end{align}


\begin{thebibliography}{10}

\bibitem{Note_QKD}
Check out http://www.magiqtech.com and http://www.idquantique.com.

\bibitem{Chapuran09a}
T.~E. Chapuran, {\em et al.}, New Journal of Physics {\bf 11,} 105001 (2009).

\bibitem{Briegel98a}
H.-J. Briegel, W. D\"ur, J. I. Cirac, and P. Zoller,
Phys. Rev. Lett. {\bf 81,} 5932 (1998).

\bibitem{Razavi09a}
M. Razavi, M. Piani, and N. L\"utkenhaus, Phys. Rev. A {\bf 80,} 032301 (2009).

\bibitem{SPIE}
M. Razavi, K. Thompson, H. Farmanbar, M. Piani, and N. L\"utkenhaus, Proc. SPIE {\bf 7236}-03, San Jose, CA (2009).

\bibitem{OFC}
M. Razavi, H. Farmanbar, and N. L\"utkenhaus, 
OFC'08 Technical Digest, Paper JWA48, San Diego, CA (2008).

\bibitem{ike_review}
D.R. Leibrandt, {\em et al.}, Quant. Inf. Comp. {\bf 9,} 0901 (2009).

\bibitem{sussex}
M. Keller, B. Lange, K. Hayasaka, W. Lange, and H. Walther, Journal of Modern Optics {\bf 54,} 1607 (2007).

\bibitem{DLCZ}
L.~M. Duan, M.~D. Lukin, J.~I. Cirac, and P.~Zoller,
{Nature} {\bf 414,} 413 (2001).

\bibitem{DLCZvariants}
B. Zhao {\em et al.}, Phys. Rev. Lett. {\bf 98,} 240502 (2007); L. Jiang {\em et al.}, Phys. Rev. A {\bf 76,} 012301 (2007); N. Sangouard {\em et al.}, Phys. Rev. A {\bf 77,} 062301 (2008); M. Gao {\em et al.}, Phys. Rev. A {\bf 79,} 042301 (2009).

\bibitem{Razavi06}
M.~Razavi and J.~H. Shapiro,
 Phys. Rev. A {\bf 73,} 042303 (2006).

\bibitem{kimble_DLCZ} C. W. Chou, J. Laurat, H. Deng, K. S. Choi, H. de Riematten, D. Felinto, H. J. Kimble,
Science {\bf 316,} 1316 (2007).

\bibitem{cohtime}
Bo Zhao, Yu-Ao Chen, Xiao-Hui Bao, Thorsten Strassel, Chih-Sung Chuu, Xian-Min Jin, Jörg Schmiedmayer, Zhen-Sheng Yuan, Shuai Chen, and Jian-Wei Pan, Nature Physics {\bf 5,} 95 (2008); R. Zhao, Y. O. Dudin, S. D. Jenkins, C. J. Campbell, D. N. Matsukevich, T. A. B. Kennedy, and A. Kuzmich, Nature Physics {\bf 5,} 100 (2008).

\bibitem{BB84} 
C. H. Bennett, Phys. Rev. Lett. {\bf 68,} 3121 (1992).

\bibitem{Albota06a}
Marius A. Albota, Franco N. C. Wong, and Jeffrey H. Shapiro, J. Opt. Soc. Am. B {\bf 23,} 918 (2006).

\bibitem{preskill_shor}
P. W. Shor and J. Preskill, Phys. Rev. Lett. {\bf 85,} 441 (2000)

\bibitem{Squash}
N. J. Beaudry, T. Moroder, and N. L\"utkenhaus, Phys. Rev. Lett. {\bf 101,} 093601 (2008); T. Tsurumaru and K. Tamaki, Phys. Rev. A {\bf 78,} 032302 (2008).

\bibitem{TobiNJP}
T. Moroder, M. Curty, and N. L\"utkenhaus, New J. Phys. {\bf 11,} 045008 (2009).

\bibitem{QKDbiased}
H.-K. Lo, H. F. Chau, and M. Ardehali, J. of Cryptology {\bf 18,} 133 (2005).








\end{thebibliography}
\end{document}